\documentclass[3p,twocolumn]{elsarticle}
\pdfoutput=0 
\usepackage{graphicx}
\usepackage{amsfonts}
\usepackage{amsmath}
\usepackage{amssymb}
\usepackage{url}
\usepackage{color}
\usepackage{epstopdf}

\def\r{{\bf r}}

\def\bunit{{\textrm{ms}/\mu\textrm{m}^2}}
\def\dunit{{\mu\textrm{m}^2/\textrm{ms}}}
\def\lunit{{\mu\textrm{m}}}

\journal{Microscopic and Mesoscopic Materials}

\begin{document}

\begin{frontmatter}
\title{Diffusion MRI/NMR at high gradients: challenges and perspectives}

\author{Denis~S.~Grebenkov\fnref{ack}}
\address{
Laboratoire de Physique de la Mati\`{e}re Condens\'{e}e, \\ 
CNRS -- Ecole Polytechnique, University Paris-Saclay, 91128 Palaiseau, France}
\ead{denis.grebenkov@polytechnique.edu}

\fntext[ack]{
The author acknowledges partial support by the grant
ANR-13-JSV5-0006-01 of the French National Research Agency.  We also
acknowledge INRIA for providing a private copy of the simulation code,
developed by Jing-Rebecca Li and Dang Van Nguyen of Equipe DEFI,
INRIA-Saclay, that was used to generate the numerical results.}

\date{\today}

\begin{abstract}
We discuss some challenges and recent advances in understanding the
macroscopic signal formation at high non-narrow magnetic field
gradients at which both the narrow pulse and the Gaussian phase
approximations fail.  The transverse magnetization and the resulting
signal are fully determined by the spectral properties of the
non-selfadjoint Bloch-Torrey operator which incorporates the
microstructure of a sample through boundary conditions.  Since the
spectrum of this operator is known to be discrete for isolated pores,
the signal can be decomposed onto the eigenmodes of the operator that
yields the stretched-exponential decay at high gradients and long
times.  Moreover, the eigenmodes are localized near specific boundary
points that makes the signal more sensitive to the boundaries and thus
opens new ways of probing the microstructure.  We argue that this
behavior is much more general than earlier believed, and should also
be valid for unbounded and multi-compartmental domains.  In
particular, the signal from the extracellular space is not Gaussian at
high gradients, in contrast to the common assumption of standard
fitting models.
\end{abstract}

\end{frontmatter}

Sixty years ago, Torrey has formulated the so-called Bloch-Torrey
equation that describes the evolution of the transverse magnetization
$m(\r,t)$ from an initial uniform state after an exciting $90^\circ$
rf pulse \cite{Torrey56}:
\begin{equation}
\label{eq:BT}
\frac{\partial}{\partial t} m(\r,t) = \bigl[D \nabla^2 - i \gamma (\vec G(t) \cdot \vec r) \bigr] m(\r,t) ,
\end{equation}
where $D$ is the diffusion coefficient, $\nabla^2$ the Laplace
operator, $i$ the imaginary unit, $\gamma$ the gyromagnetic ratio, and
$\vec G(t)$ the time-dependent gradient (that also includes the effect
of refocusing rf pulses).  The MR signal $S$ at time $t$ is obtained
by integrating $m(\r,t)$ over a voxel or the whole sample.  The
microstructure is incorporated {\it implicitly} through appropriate
boundary conditions that describe the behavior of the nuclei at the
boundaries (e.g., impenetrable walls, surface relaxation or diffusive
exchange between adjacent compartments).  From the mathematical point
of view, the two major challenges of this linear diffusion equation
are the imaginary character of the encoding term (yielding a
non-self-adjoint governing operator) and boundary conditions.  As a
consequence, this equation does not admit simple explicit solutions
except for free diffusion (i.e., without microstructure), for which
\begin{equation}
\label{eq:free}
S = S_0 \exp\bigl(- bD \bigr),
\end{equation}
where $S_0$ is the reference signal without gradient (that also
incorporates $T_1$ and $T_2$ relaxation that we excluded from
Eq. (\ref{eq:BT})), and $b$ is related to the gradient $\vec G(t)$
\cite{Stejskal65}.

Since the seminal paper by Torrey, numerous theoretical approaches
have been developed to relate the microstructure of a studied sample
(e.g., a tissue or an oil-bearing rock) to the macroscopic signal and
then to infer some structural information about the sample from the
measured signal (see \cite{Price,Grebenkov07} and references therein).
For instance, in the narrow pulse approximation, the assumption of
infinitely narrow gradient pulses allows one to effectively remove the
encoding term in Eq. (\ref{eq:BT}) and thus to reduce the problem to
pure diffusion.  In turn, the Gaussian phase approximation treats the
encoding term perturbatively that leads to an appropriate modification
of the free diffusion signal in Eq. (\ref{eq:free}) at weak gradients.
While both approximations are successfully used to interpret measured
signals in various applications, they remain limited by the underlying
assumptions.  In particular, there are many experimental evidences of
the non-Gaussian signal decay \cite{Niendorf96,Jensen05,Trampel06}
that are typically attributed to (i) pore size distribution resulting
in a superposition of Gaussian signals,
the most famous example being the bi-exponential model for the MR
signal from intracellular and extracellular regions;
or (ii) exchange between compartments \cite{Karger88}.  Overall, most
of theoretical works aimed at keeping the Gaussian hypothesis at any
cost, sometimes in a disguised form \cite{Grebenkov10b}.

At the same time, the theoretical foundation of the Gaussian
hypothesis has been broken 25 years ago by Stoller, Happer and Dyson
\cite{Stoller91} who provided an exact solution of the Bloch-Torrey
equation for two simple one-dimensional configurations: the semi-axis
$(0,\infty)$ and an interval $(0,L)$, both with reflecting endpoints.
They showed that the macroscopic signal exhibits nontrivial
stretched-exponential decay at high non-narrow gradients:
\begin{equation}
\label{eq:stretched}
\ln (S/S_0) \propto G^{\frac 23} \propto b^{\frac 13}.
\end{equation}
This behavior drastically differs from the conventional Gaussian decay
in Eq. (\ref{eq:free}).  Three years later, de Swiet and Sen derived
the same asymptotic behavior for impermeable disk and sphere and
argued that it should be generic for restricted diffusion in isolated
bounded domains \cite{deSwiet94}.  H\"urlimann {\it et al.} confirmed
these theoretical predictions in a pulsed-gradient spin-echo (PGSE)
experiment for water diffusion between two parallel plates
\cite{Hurlimann95}.  The measured signal was shown to decay as
Gaussian at small gradients and then to switch to the above
stretched-exponential decay at higher gradients.  Remarkably, the
transition between two regimes occurred at a moderate gradient of
$15$~mT/m, available at any modern clinical MR scanner.  Since these
pioneering works, no significant progress has been achieved in
understanding the signal formation at high gradients, in spite the
fact that very high gradients, up to few T/m, are available nowadays
and become more and more often used in experiments to achieve higher
sensitivity and selectivity (e.g., $b$-values up to $40~\bunit$, or
$40\,000~$s/mm$^2$, were used for brain imaging \cite{Wedeen12}).
Such an ignorance to this field from the theoretical MR community can
partly be explained by: (i) rather involved, complicated mathematics
needed to analyze the Bloch-Torrey equation at high gradients; and
(ii) too small signals, often at the noise level, that makes
challenging their practical use.  Even if the purposeful application
of high gradients may still be problematic nowadays (partly because of
the lack of an appropriate theory to interpret such measurements), the
outcomes of current experiments can be drastically affected by
unnoticed, unexplained or misleadingly interpreted deviations from the
Gaussian behavior.  In this paper, we show that the
stretched-exponential decay of the signal is a generic, universal
feature of diffusion NMR that can be the origin of the non-Gaussian
behavior in many experiments.

For the sake of clarity, we start with the time-independent gradient
profile, whereas the results for spin echoes after rectangular
gradient pulses are obtained by combining solutions of the above
equation in a standard way (see \cite{Grebenkov07} for details).  In
this case, the solution of the Bloch-Torrey equation (\ref{eq:BT}) can
be formally written as
\begin{equation}
m(\r,t) = \exp\biggl(- \underbrace{\bigl[-\nabla^2 + i g x\bigr]}_{\rm BT-operator} Dt\biggr) m_0 ,
\end{equation}
where $g = \gamma G/D$, $m_0$ is the initial magnetization, and $x$ is
the coordinate along which the gradient is applied.  We call the
expression in square parentheses the Bloch-Torrey (or BT) operator
which englobes the whole complexity of the problem (in particular, the
microstructure).  The transverse magnetization and the resulting MR
signal are thus fully determined by the spectral properties of this
non-selfadjoint operator.

For any bounded domain (e.g., an isolated pore), the spectrum of the
BT-operator is discrete and bounded from below (it follows from the
discrete spectrum of the Laplace operator and boundness of the term
$igx$).  As a consequence, the MR signal can be written as a spectral
decomposition over eigenmodes of the BT-operator:
\begin{equation}
\label{eq:S}
S = \sum\limits_n A_n(g) \, \exp\bigl(- \lambda_n(g)\, Dt \bigr) ,
\end{equation}
where $\lambda_n(g)$ are the eigenvalues and $A_n(g)$ are the squared
projections of the corresponding eigenfunctions of the BT-operator
onto a constant function.  This representation, which is similar to
the famous Brownstein-Tarr expansion derived for a simpler diffusion
problem without gradients \cite{Brownstein79}, is general and valid
for any bounded domain, gradient amplitude and time.  In particular,
the eigenvalue with the smallest real part determines the signal decay
at long times.  The asymptotic behavior of the eigenvalues
$\lambda_n(g)$ at high gradients ($g\to\infty$) was first analyzed for
an interval \cite{Stoller91} and then for a disk and a sphere
\cite{deSwiet94}, all with reflecting boundaries.  The obtained
leading behavior $\lambda_n(g) \propto g^{\frac23}$ implied the
stretched-exponential decay in Eq. (\ref{eq:stretched}).

More recently, the analysis was extended to multiple intervals with
semi-permeable boundaries \cite{Grebenkov14b,Grebenkov16a} and to
arbitrary planar domains \cite{Grebenkov16b}.  In the former papers,
the effect of diffusive exchange between compartments onto the MR
signal was investigated (see also \cite{Grebenkov14a} for a recent
overview and references on this topic).  In turn, a general asymptotic
construction presented in \cite{Grebenkov16b} extends and improves the
earlier result by de Swiet and Sen for bounded domains.  It was shown
that, for large enough $g$, the eigenfunctions of the BT-operator are
localized near the boundary points $\r_j$ at which the normal vector
to the boundary is parallel to the gradient direction.  The four-term
asymptotic behavior of the corresponding eigenvalues in powers of
$g^{\frac16}$ was obtained, the leading term $g^{\frac23}$ being
universal.  In turn, boundary curvature, surface relaxation and
membrane permeability were shown to affect the sub-leading terms of
order $g^{\frac12}$ and $g^{\frac 13}$.  For a PGSE experiment
with rectangular gradient pulses, the representation (\ref{eq:S}) can
be adapted \cite{Grebenkov14b,Grebenkov16b}.

\begin{figure}
\begin{center}
\includegraphics[width=78mm]{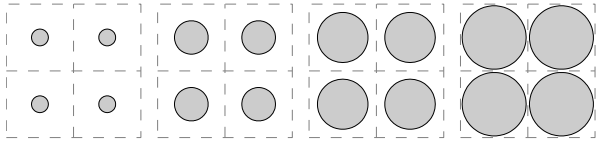} 
\end{center}
\caption{
Hindered diffusion in the exterior (white) space of a periodic
arrangment of shadowed circular obstacles (cylinders) of radius $R$
separated by distance $2L$.  Four configurations are shown for $L =
20~\mu\rm{m}$ and $R = 5~\mu\rm{m},\, 10~\mu\rm{m},\, 15~\mu\rm{m},\,
19~\mu\rm{m}$.  Dashed lines indicate one computational domain which
is repeated periodically in both directions.  The volume fraction of
fluid phase (white region), $1 - \pi R^2/(2L)^2$, is respectively
$0.95$, $0.80$, $0.56$ and $0.29$.}
\label{fig:domain}
\end{figure}

Most importantly, the above asymptotic analysis is local and therefore
independent of whether the domain is bounded or not.  As a
consequence, the localization of eigenfunctions and the asymptotics of
eigenvalues as $g\to\infty$ are conjectured to hold for unbounded
domains as well.  For instance, we conjecture that the Bloch-Torrey
operator for the exterior of a disk or a sphere has a discrete
spectrum.  Although its mathematical proof is still missing, it is
strongly supported by numerical results \cite{Grebenkov16b}.  This is
an important statement that may drastically change the current view
onto diffusion NMR.  In fact, the gradient encoding term, $igx$, was
always considered as a perturbation of the Laplace operator
$\nabla^2$, yielding the conventional Gaussian paradigm.  This was
mathematically justified for isolated pores.  In turn, while the
Laplace operator has a continuous spectrum for unbounded domains, the
inclusion of the term $igx$, even with arbitrarily small $g$, is
expected to make the spectrum of the BT-operator discrete so that
$igx$ cannot be considered as a perturbation anymore.  Moreover, the
limit $g\to 0$, in which the discrete spectrum should become
continuous, turns out to be singular.  If former theories relied on
the Gaussian signal (\ref{eq:free}) and considered the
stretched-exponential decay (\ref{eq:stretched}) as ``pathologic'', a
new theory has to rely on the spectral decomposition (\ref{eq:S}) as
the starting point and then explain how the Gaussian decay is
recovered at weak gradients.  Such an analysis was already performed
for the case of an interval in \cite{Grebenkov14b} but further
investigations are necessary for more realistic bounded and unbounded
domains.

What does this mathematical discussion change in practice?  For
instance, all standard models of diffusion NMR signal in brain tissue
assumed the extracellular (or extraneuronal) signal to be Gaussian
(e.g., see \cite{Assaf04}).  To check this assumption, we study
hindered diffusion of water molecules in the exterior space of an
infinite periodic configuration of impermeable circular obstacles
(cylinders) of radius $R$ whose centers form a square lattice with
spacing $2L$ (Fig. \ref{fig:domain}).  We solve the Bloch-Torrey
equation by a finite elements method \cite{Nguyen14} and then compute
the macroscopic signal for a PGSE experiment with the gradient
pulse duration $\delta$ and the inter-pulse time $\Delta$, by setting
$\delta = \Delta = 10$~ms, $D = 3~\dunit$, and $\gamma = 2.675 \cdot
10^8$~rad/T/s.  Here we vary the gradient $G$ to get a broad range of
$b$-values up to $30~\bunit$, with $b = (\gamma G)^2 \delta^2 (\Delta
- \delta/3)$.  The gradient is applied along the diagonal direction
$(1,1)$.

\begin{figure}
\begin{center}
\includegraphics[width=75mm]{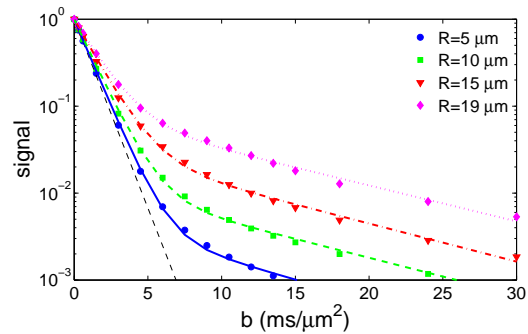} 
\end{center}
\caption{
PGSE signal as a function of $b$-value for the exterior space of a
periodic arrangment of 2D disks of radius $R$ separated by distance
$2L=40~\mu$m, with $D = 3~\dunit$, $\delta = \Delta = 10$~ms and
impermeable boundary.  Different symbols correspond to $R =
5~\lunit,\, 10~\lunit,\, 15~\lunit,\, 19~\lunit$.  Lines present
bi-exponential fits of simulated signals over the range of $b$-values
up to $10~\bunit$.  Thin dashed line shows the signal (\ref{eq:free})
for free diffusion.  The accuracy of simulations was validated by
doubling the mesh size and checking the smallness of deviations
between two cases (not shown).}
\label{fig:dang_cdisk}
%
\end{figure}

Figure \ref{fig:dang_cdisk} shows the signal $S$ versus $b$-value for
four choices of the obstacle size $R$ ranging from $5~\lunit$ to
$19~\lunit$.  One can clearly see the non-Gaussian features already 
at moderate $b$-values used in brain dMRI.  For
instance, in the case of large obstacles, deviations from the Gaussian
behavior start at $b$-values of the order of $1~\bunit$, while the
signal remains measurable (say, above $0.01$) even at very large
$b$-values.  If these signals were conventionally fitted by a
bi-exponential model, the apparent good agreement would lead to false
interpretations of the signal behavior and meaningless fitting
parameters.

While the simple microstructure of periodic circular obstacles was
chosen to facilitate numerical computation, the observed
stretched-exponential decay is expected to emerge for {\it any}
nontrivial microstructure which can be two- or three-dimensional,
bounded or unbounded, periodic or random, mono- or polydisperse, with
obstacles of any shape, and even for multi-compartmental domains with
semi-permeable interfaces.  The observed localization phenomenon is
thus generic and may be relevant for most physical and biomedical
diffusion MRI/NMR applications.  In particular, the Gaussian
assumption of the extracellular signal may not be valid in many
practical situations.

Although we presented the signal as a function of $b$-value for
convenience of a broad MRI community, it is important to emphasize
that the signal in (\ref{eq:S}) exhibits different dependences on the
gradient parameters: $G$ enters intrinsicly into spectral properties
of the BT-operator, $\delta$ ($=t$) stands only in the
exponential function, whereas $\Delta$ is included through the
coefficients $A_n$.  While the universal form of the single $b$-value
incorporating the experimental setup was helpful for the analysis at
low $b$-values, the use of this parameter as a single descriptor of a
diffusion NMR experiment is in general misleading.  The parameters
$G$, $\delta$ and $\Delta$ play different roles and affect the signal
in distinct ways.  While such non-universal dependences are more
difficult to analyze and to control, they make the diffusion NMR
richer, more sensitive and selective.

We conclude that the stretched-exponential decay of the MR signal at
high gradients is not, as commonly believed, a pathological exception
from the Gaussian realm but the intrinsic feature of the BT-operator
for almost any microstructure.  This finding can become a first step
towards a new theory of signal formation at high gradients that should
replace unconsolidated and sometimes misleading phenomenological 
fitting models used to interpret diffusion NMR measurements.

\end{document}